\documentstyle[epsfig,12pt]{article}

\def \bea{\begin{eqnarray}}
\def \beq{\begin{equation}}

\def \eea{\end{eqnarray}}
\def \eeq{\end{equation}}

\begin{document}

\begin{flushright}
DESY 02-195
\end{flushright}

\begin{center}

{\large\sc Phenomenology of Non-Commutative Field Theories}

\end{center}

\begin{center}
 
Stephen Godfrey$^{a,b}$
and 
M.A. Doncheski$^c$

\vspace*{0.3cm}

{\it $^a$  Department of Physics, Carleton University, 
Ottawa, ON K1S 5B6, Canada} \\

\vspace*{0.2cm}
{\it $^b$ DESY, Deutsches Elektronen-Synchrotron, D22603 Hamburg, 
Germany} \\

\vspace*{0.2cm}

{\it $^c$Department of Physics, Pennsylvania State University,
Mont Alto, PA 17237 USA} \\

\end{center}

\begin{abstract}
\noindent
We study the effects of non-commutative QED (NCQED) in fermion pair 
production, $\gamma + \gamma \rightarrow f + \bar{f}$ and Compton scattering, 
$e + \gamma \rightarrow e + \gamma$.  Non-commutative geometries 
give rise to 3- and 4-point 
photon vertices and to momentum dependent phase factors in QED vertices which 
will have observable effects in high energy collisions.  We consider 
$e^+ e^-$ colliders with energies appropriate to the TeV Linear Collider 
proposals and the multi-TeV CLIC project operating in $\gamma \gamma$ and 
$e\gamma$ modes.  Non-commutative scales roughly equal to the center of mass 
energy of the $e^+e^-$ collider can be probed, with the exact value depending 
on the model parameters and experimental factors.  The 
Compton process is sensitive to $\Lambda_{NC}$ values roughly twice as large 
as those accessible to the pair production process.
\end{abstract}

\section{Introduction}

Non-commutative quantum field theories (NCQFT) \cite{dn02} 
arise in string/M 
theory by describing the low-energy excitations of D-branes in a 
background of ``EM-like'' fields.  Numerous ideas originating in 
string theory have stimulated particle physics phenomenology.  We can 
regard NCQFT as yet another such example.  NCQFT gives rise to an 
interesting set 
of operators not yet explored and therefore not ruled out.  These give 
rise to a rich phenomenology with 
testable differences between QFT and NCQFT \cite{hk02}.  In this 
contribution we give a brief overview of NCQFT phenomenology with an 
emphasis on limits that can be extracted at high energy $e^+e^-$ 
colliders \cite{gd02,hpr}.

\section{Non-commutative Field Theory}

The central idea of NCQFT is that the conventional commuting 
coordinates are replaced with non-commuting space-time operators:
\begin{equation}
[\hat{X}_\mu, \hat{X}_\nu] = i\theta_{\mu\nu} \equiv {i\over 
{\Lambda_{NC}^2}} C_{\mu\nu}
\end{equation}
Here we adopt the Hewett-Petriello-Rizzo parametrization \cite{hpr}
where the 
overall scale, $\Lambda_{NC}$, characterizes the threshold where 
non-commutative (NC) effects become relevant and 
$C_{\mu\nu}$ is a real antisymmetric 
matrix  whose dimensionless elements are presumably of 
order unity.  One might expect the scale $\Lambda_{NC}$ to be of order 
the Planck scale.  However, given the possibility of large extra 
dimensions \cite{add,rs} 
where gravity becomes strong at scales of order a TeV, it 
is possible that NC effects could set in at a TeV.  We therefore 
consider the possibility that $\Lambda_{NC}$ may lie not too far above 
the TeV scale.

We stress that 
the $C$ matrix is not a tensor since its elements are identical in all 
reference frames resulting in the violation of Lorentz invariance. 
Thus,  NCQFT violates Lorentz invariance at $\Lambda_{NC}$.
The $C$-matrix
can be parameterized, following the notation of \cite{JoA}, as
\begin{equation}
C_{\mu \nu} = \left(
\begin{array}{cccc}
0 & C_{01} & C_{02} & C_{03} \\
-C_{01} & 0 & C_{12} & -C_{13} \\
-C_{02} & -C_{12} & 0 & C_{23} \\
-C_{03} & C_{13} & -C_{23} & 0 \\
\end{array}
\right)
\end{equation}
where $\sum_i |C_{0i}|^2 = 1$.  
Thus, the $C_{0i}$ are related to space-time NC and 
are defined by the direction of the background ``{\bf E}-field''.  
Furthermore, the $C_{0i}$ can be parameterized as 
\begin{eqnarray}
C_{01} & = & \sin \alpha \cos \beta \nonumber \\
C_{02} & = & \sin \alpha \sin \beta \nonumber \\
C_{03} & = & \cos \alpha.
\end{eqnarray}
Likewise, the $C_{ij}$ are related 
to the space-space non-commutativeness and are defined by the direction of the 
background ``{\bf B}-field''.  They can be parameterized as 
\begin{eqnarray}
C_{12} & = & \cos \gamma \nonumber \\
C_{13} & = & \sin \gamma \sin \beta \nonumber \\
C_{23} & = & -\sin \gamma \cos \beta.
\end{eqnarray}
In both cases 
$\beta$ defines the origin of the $\phi$ axis which we set 
to $\beta = \pi/2$ so that we can parametrize $C$ with the two 
angles $\alpha$ and $\gamma$, the angles of the 
background ``{\bf E}-field''  and ``{\bf B}-field''
relative to the $z$-axis.  
Since experiments are sensitive to the direction of the $C$-vectors an 
astronomical coordinate systems must be employed and events must be 
time stamped event by event.  

NCQFT can be cast in the from of a conventional QFT using two 
approaches: the Seiberg-Witten mapping \cite{sw} and the Weyl-Moyal 
approach \cite{wm}.
In the Moyal-Weyl approach only $U(N)$ Lie algegras are closed under 
Moyal brackets so that NC gauge theories are only based on $U(N)$ 
groups and covariant derivatives can only be constructed for fields 
with $Q=0$, $\pm 1$.  In contrast a NCSM has been developed using the 
Sieberg-Witten approach but it is non-renormalizable order by order.

It is difficult to construct the NCSM but NCQED is well defined in the 
Moyal-Weyl approach.  It is the simplest extension of the standard 
model so that we will
focus our
discussion on NCQED but note that others have studied the NCSM.  This 
approach provides a testing ground for the ideas of NCQFT.

To study the phenomenology of NCQED non-commuting coordinates 
are introduced and the Lagrangian is then cast in terms of 
conventional commuting quantum fields \cite{ihab}.  
This gives rise to several 
modifications of the standard commuting QED.  It introduces momentum 
dependent phase factors in the fermion-photon vertices and introduces 
new three and four-point photon vertices.  These changes violate 
Lorentz invariance.  The hallmark signal is azimuthal dependencies in 
the cross sections in $2\to 2$ processes resulting from the existence 
of a preferred direction.

We note another approach to non-commuting phenomenology is to consider 
Lorentz violating operators such as
\begin{equation}
\theta^{\mu\nu} \bar{q} \sigma_{\mu\nu} q
\end{equation}
to place limits on the scale noncommutativity \cite{ccl,mpr}.  The 
$\theta^{\mu\nu} \bar{q} \sigma_{\mu\nu} q$ operator acts like a 
$\vec{\sigma}\cdot \vec{B}$ interaction with $\vec{B}$ fixed. 
The different sensitivities of Cs and Hg atomic clocks to external 
$\vec{\sigma}\cdot \vec{B}$ leads to the estimate of \cite{ccl}
$\Lambda_{NC}>10^{17}$~GeV.  The operator 
$\theta^{\mu\nu} \bar{N} \sigma_{\mu\nu} N$, where $N$ are nucleon 
wavefunctions, leads to a shift in nuclear magnetic moments.  
Measurements of sidereal variation in hyperfine splittings in atoms 
gives $\Lambda_{NC}>10^{15}$~GeV \cite{mpr}.  
If one accepts these conclusions it 
is unlikely that colliders can probe NC phenomenology.  The standard 
way out is the supposition that loop contributions might be cancelled by other 
new physics. Some loop induced tests of 
non-commutative field theories are summarized in Table 1.

\begin{table}[h]
\begin{center}
Table 1: Loop induced tests of NCSM.
\vskip0.2cm
\begin{tabular}{|l|l|}
\hline
Process 	& Reference \\
\hline
$b\to s\gamma, \; sg$ & Iltan \cite{iltan} \\
Lamb shift & Chaichian, Sheikh-Jabbari and Tureanu \cite{cst} \\
CP violation & Chang and Xing \cite{cx} \\
	& Hinchliffe and Kersting \cite{hk} \\
$(g-2)_\mu$ & Kersting \cite{kersting} \\
Hyperfine Structure & Mocioiu, Pospelov and Roiban \cite{mpr} \\
$Z\to \gamma\gamma, \; gg$ & Mocioiu, Pospelov and Roiban \cite{mpr} \\
	& Behr {\it et al.} \cite{behr} \\
$\pi \to \gamma\gamma\gamma$ & Grosse and Liao \cite{gl} \\
$Z\to \ell^+ \ell^-$, $W\to \ell \nu$ & Iltan \cite{iltanb} \\
\hline
\end{tabular}
\end{center}
\end{table}

\section{Collider Tests of NCQED}

Collider tests of non-commutative field theories have mainly 
concentrated on NCQED and by implication $e^+e^-$ colliders.  In this 
context a number of processes have been studied: Compton scattering, 
$e\gamma\to e\gamma$ \cite{gd02,mathews}, 
pair production, $\gamma\gamma \to e^+e^-$
\cite{gd02,baek}, 
pair annihilation $e^+e^-\to \gamma\gamma$ \cite{hpr}, Moller 
scattering, 
$e^-e^-\to e^-e^-$  \cite{hpr}, Bhabha scattering, $e^-e^+\to e^-e^+$  \cite{hpr}, 
two photon 
scattering, $\gamma\gamma\to \gamma\gamma$  \cite{hpr}, and Higgs production, 
$\gamma\gamma \to H^+ H^-$ \cite{gl01}.  

We will concentrate on the two processes $e\gamma\to e\gamma$ and 
$\gamma\gamma \to e^+e^-$ \cite{gd02} as they provide good examples of the 
behavior we expect in $2\to 2$ processes in general.  We mention 
other processes \cite{hpr} 
where the analysis provides additional ingredients 
beyond what was done in these analysis.  We consider $\sqrt{s}=0.5$, 
0.8, 1.0, 1.5, 3, 5, and 8~TeV appropriate to the TESLA/LC/CLIC 
proposals.  We assume an integrated luminosity of $L=500$~fb$^{-1}$.  
To take into account finite detector acceptance we assume angular 
acceptance of $10^o\leq \theta \leq 170^o$ and $p_T>10$~GeV.

\subsection{Pair Production: $\gamma\gamma\to e^+e^-$}

The Feynman diagrams for $\gamma\gamma\to e^+e^-$ are shown in Fig. 
1.  There are two important differences between the non-commuting 
version of this process and SM QED.  The first is that there are 
momentum dependent phases in the fermion-photon vertices and the 
second is the introduction of the third diagram which includes a 
tri-linear photon  vertex.  The resulting cross section is given by
\cite{gd02}:
\begin{equation}
{{d\sigma(\gamma\gamma\to f\bar{f})}\over{d\cos\theta \; d\phi}} = {{\alpha^2}\over{2s}}
\left\{ \frac{\hat{u}}{\hat{t}} + \frac{\hat{t}}{\hat{u}} - 
4 \frac{\hat{t}^2 + \hat{u}^2}{\hat{s}^2} \sin^2 
\left(\frac{k_1 \cdot \theta \cdot k_2}{2} \right) \right\}
\end{equation}
where $k_1$ and $k_2$ are the momenta of the incoming photons. The 
first two terms in the cross section are the standard model result.  
The third term is the NCQED contribution which includes contributions 
from the fermion-photon phase factors as well as the diagram with the 
tri-linear photon vertex.
The bilinear product in the phase factor is given by:
\begin{equation}
\frac{1}{2} k_1 \cdot \theta \cdot k_2 = \frac{\hat{s}}{4 \Lambda_{NC}^2} 
C_{03}= \frac{\hat{s}}{4 \Lambda_{NC}^2}  \cos\alpha.
\end{equation}
This phase factor breaks Lorentz invariance.  There is no $\phi$ 
dependence in this cross section and as $\Lambda_{NC}\to \infty$ the 
phase angle goes to 0 and the SM is recovered.  

\begin{figure}[t]
\begin{center}
\includegraphics[bb=22 458 377 674,width=3.5cm,clip=true]{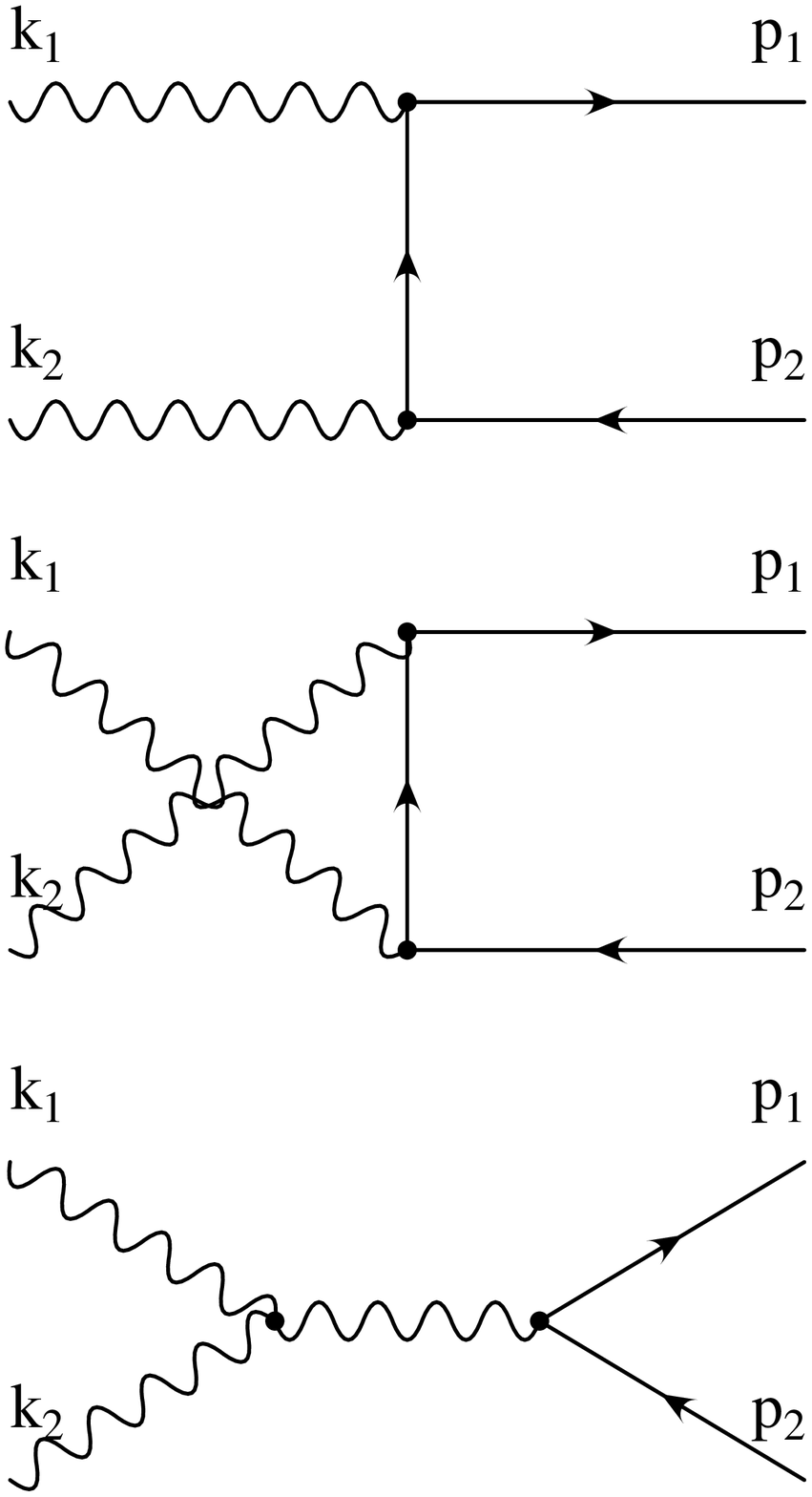} 
$\qquad$
\includegraphics[bb=22 226 377 460,width=3.5cm,clip=true]{feynd_pair.eps} 
$\qquad$
\includegraphics[bb=22 0 377 220,width=3.5cm,clip=true]{feynd_pair.eps} 
\caption{The Feynman diagrams contributing to the process 
$\gamma \gamma \to e^+ e^-$.}
\end{center}
\end{figure}

Fig.~2 shows the cross section for $\gamma\gamma\to e^+e^-$ vs. $\Lambda_{NC}$ 
for QED and NCQED with $\alpha = 0$ and $\pi/4$, for a $\sqrt{s} = 0.5$~TeV 
$e^+ e^-$ collider operating in $\gamma \gamma$ mode
\cite{backlaser}.  The event rate is high 
with statistics that can exclude NCQED to a fairly high value of 
$\Lambda_{NC}$.  Note that the QED (solid) curve is actually a central QED 
value with $\pm 1 \sigma$ bands (assuming 500~fb$^{-1}$ of integrated 
luminosity).  

\begin{figure}[t]
\begin{center}
\rotatebox{-90}{
\includegraphics[width=6.0cm,clip=true]{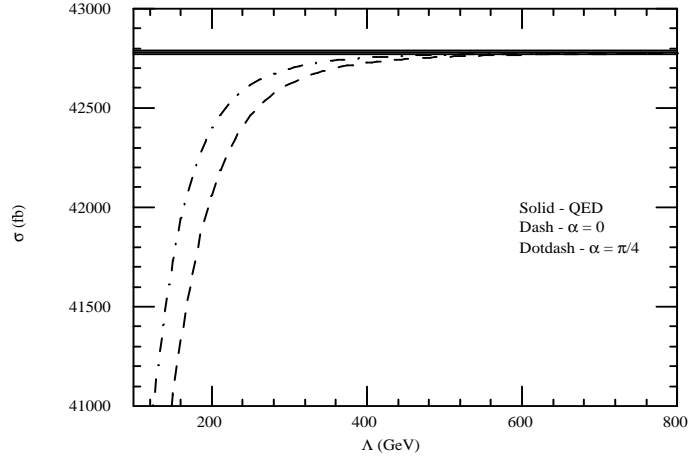} 
}
\caption{$\sigma$ vs. $\Lambda_{NC}$ for the pair production process, 
$\sqrt{s} = 500$ GeV.  The solid line corresponds to the SM cross section 
$\pm$ 1 standard deviation (statistical) error 
based on integrated luminosity of $L = 500 fb^{-1}$.}
\end{center}
\end{figure}

To quantify the sensitivity to NCQED, we calculate the $\chi^2$ for 
the deviations between NCQED and the SM for a range  of parameter values.  We 
start by calculating statistical errors based on an integrated luminosity of 
500~fb$^{-1}$.  We consider two 
possibilities for systematic errors.  In the first case we do not include 
systematic errors while in the second case we obtain limits by combining a 2\% 
systematic error combined in quadrature with the statistical errors; 
$\delta =\sqrt{\delta^2_{stat} + \delta^2_{sys}}$.  The 2\% systematic error 
is a very conservative estimate of systematic errors, for example the TESLA 
TDR calls for only a 1\% systematic error.  
We calculate the $\chi^2$ for the following obervables; total cross 
sections, $\cos \theta$ and $\phi$ angular distributions,
binning the angular distributions into 20 bins in $\cos \theta$ and 
$\phi$.  
\begin{equation}
\chi^2_{\cal O} (\Lambda) 
=   \sum_i
\left( { \frac{ {\cal O}_i (\Lambda) - {\cal O}_i^{QED} }
{\delta {\cal O}_i } }\right) ^2 
\end{equation}
where $\cal O$ represents the observable under consideration and the sum is 
over the bins of the angular distributions.  $\chi^2 = 4$ represents a 
95\% C.L. deviation from QED, which we'll define as the sensitivity limit.
The 
$\cos \theta$ distribution consistently gives the highest exclusion limits on 
$\Lambda_{NC}$, regardless of $\sqrt{s}$ and $\alpha$ (as long as 
$\alpha \neq \pi/2$, where, again, no limits are possible). 
We obtain limits of $\Lambda_{NC}>(0.5-1)\sqrt{s}$.

\subsection{Compton scattering}

The Feyman diagrams contributing to Compton scattering are shown in 
Fig. 3.  The differential cross section is given by \cite{gd02}:
\begin{equation}
{{d\sigma(e^-\gamma\to e^-\gamma)}\over{d\cos\theta \; d\phi}} = {{\alpha^2}\over{2s}}
\left\{ -\frac{\hat{u}}{\hat{s}} - \frac{\hat{s}}{\hat{u}} + 
4 \frac{\hat{s}^2 + \hat{u}^2}{\hat{t}^2} \sin^2 
\left(\frac{k_1 \cdot \theta \cdot k_2}{2} \right) \right\}.
\end{equation}
The first two terms in the expression are the standard QED contribution, 
while the last term is due to NCQED effects.
As before, the NC phase factor only appears in this new term.

\begin{figure}[t]
\begin{center}
\includegraphics[bb=22 477 387 667,width=3.5cm,clip=true]{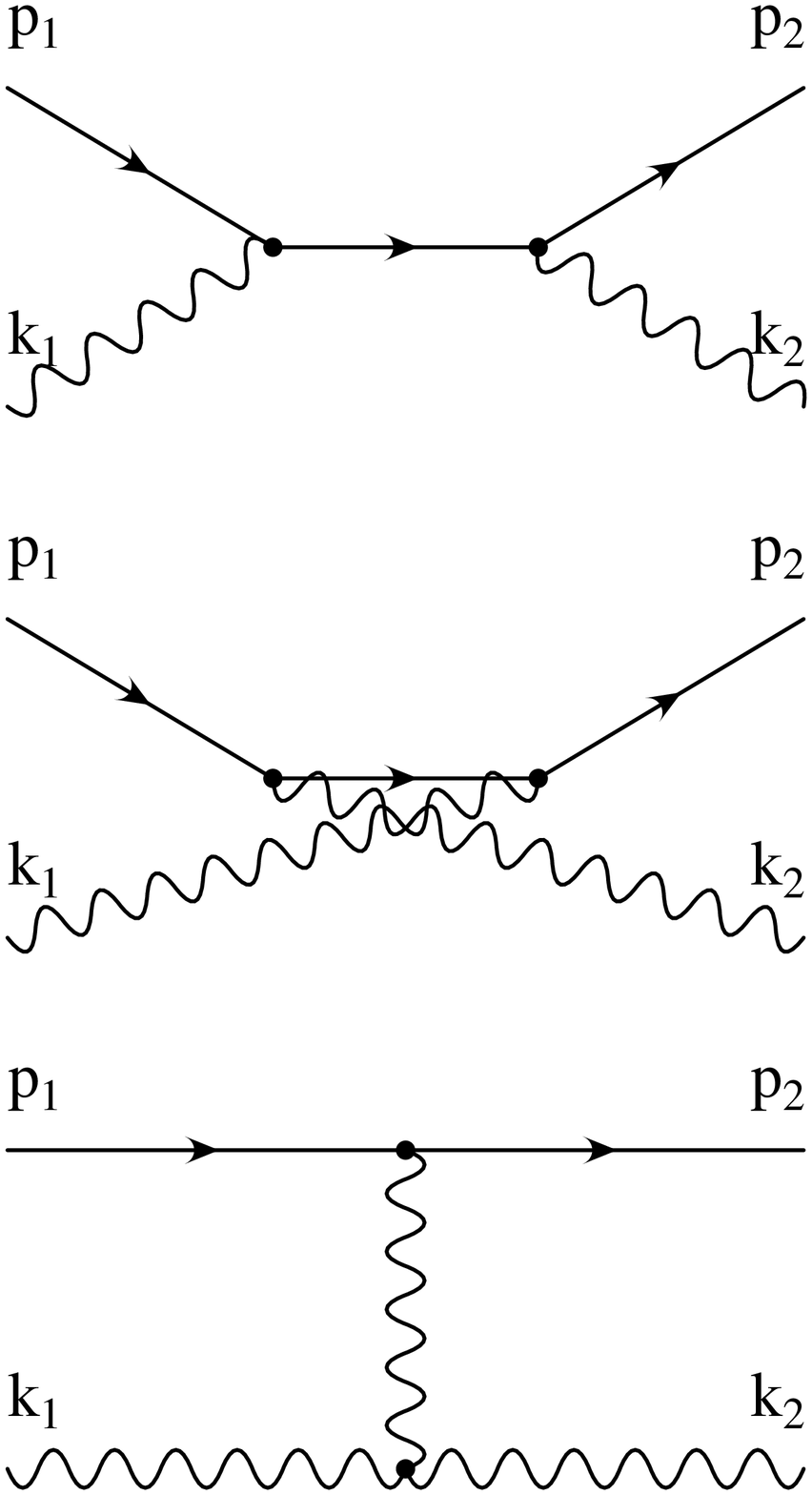} 
$\qquad$
\includegraphics[bb=22 246 378 433,width=3.5cm,clip=true]{feynd_comp.eps} 
$\qquad$
\includegraphics[bb=22 12 378 199,width=3.5cm,clip=true]{feynd_comp.eps} 
\caption{The Feynman diagrams contributing to the process 
$e \gamma \to e \gamma$.}
\end{center}
\end{figure}

Here, $p_1$ and $k_1$ are the momenta of the initial state electron and 
photon, respectively, while $p_2$ and $k_2$ are the momenta of the final state 
electron and photon, respectively.  $\hat{s}$, $\hat{t}$ and $\hat{u}$ are the 
usual Mandelstam variables $\hat{s} = (p_1 + k_1)^2$, 
$\hat{t} = (p_1 - p_2)^2$ and $\hat{u} = (p_1 - k_2)^2$.  Choosing 
$k_1 = x \frac{\mbox{$\sqrt{s}$}}{\mbox{$2$}} (1,0,0,-1)$ and 
$k_2 = k(1,\sin \theta \cos \phi, \sin \theta \sin \phi, \cos \theta)$, 
results in the following expression for the phase factor:
\begin{eqnarray}
\frac{1}{2} k_1 \cdot \theta \cdot k_2 &  = &
 \frac{x k \sqrt{s}}{4 \Lambda_{NC}^2} 
[ (C_{01} - C_{13}) \sin \theta \cos \phi + 
(C_{02} + C_{23}) \sin \theta \sin \phi + C_{03}(1 + \cos \theta) ] 
\nonumber \\
& = & \frac{x k \sqrt{s}}{4 \Lambda_{NC}^2} 
[ - \sin \gamma \sin \theta \cos \phi + 
\sin \alpha \sin \theta \sin \phi + \cos \alpha (1 + \cos \theta) ].
\end{eqnarray}
where $x$ is the momentum fraction of the incident photon, $k$ is the 
magnitude of the 3-momentum of the final state photon, and $\theta$ and $\phi$ 
are the lab frame angles of the final state photon.  
Choosing $\beta = \pi/2$ leaves us with two free 
parameters in addition to $\Lambda_{NC}$.
Compton scattering is sensitive to both space-space and 
space-time NC parts, probing $\gamma$ in addition  to $\alpha$.  It 
therefore complements $\gamma\gamma\to e^+e^-$.

Fig.~4 shows the cross section $\sigma$ vs. $\Lambda_{NC}$ for QED and NCQED 
with $\gamma = 0$ and $\pi/2$ and $\alpha = 0$, $\pi/4$ and $\pi/2$,
for a $\sqrt{s} = 0.5 \; TeV$ 
$e^+ e^-$ collider operating in $e \gamma$ mode.  The event rate is high, so 
there are enough statistics to probe NCQED up to a fairly high value of 
$\Lambda_{NC}$.  Again, the QED (solid) curve includes the central QED value 
and $\pm 1 \sigma$ bands (assuming 500~fb$^{-1}$ of integrated luminosity).  
Fig.~5 shows the angular distributions
$d\sigma/d\phi$ for QED and NCQED with $\gamma = 0$ and $\pi/2$ and
$\alpha = \pi/2$, and 
$\sqrt{s}$ = $\Lambda_{NC}$ = $500 \; GeV$.  The error bars in Fig.~5 assume 
500~fb$^{-1}$ of integrated luminosity.

\begin{figure}[h]
\begin{center}
\rotatebox{-90}{
\includegraphics[width=5.0cm,clip=true]{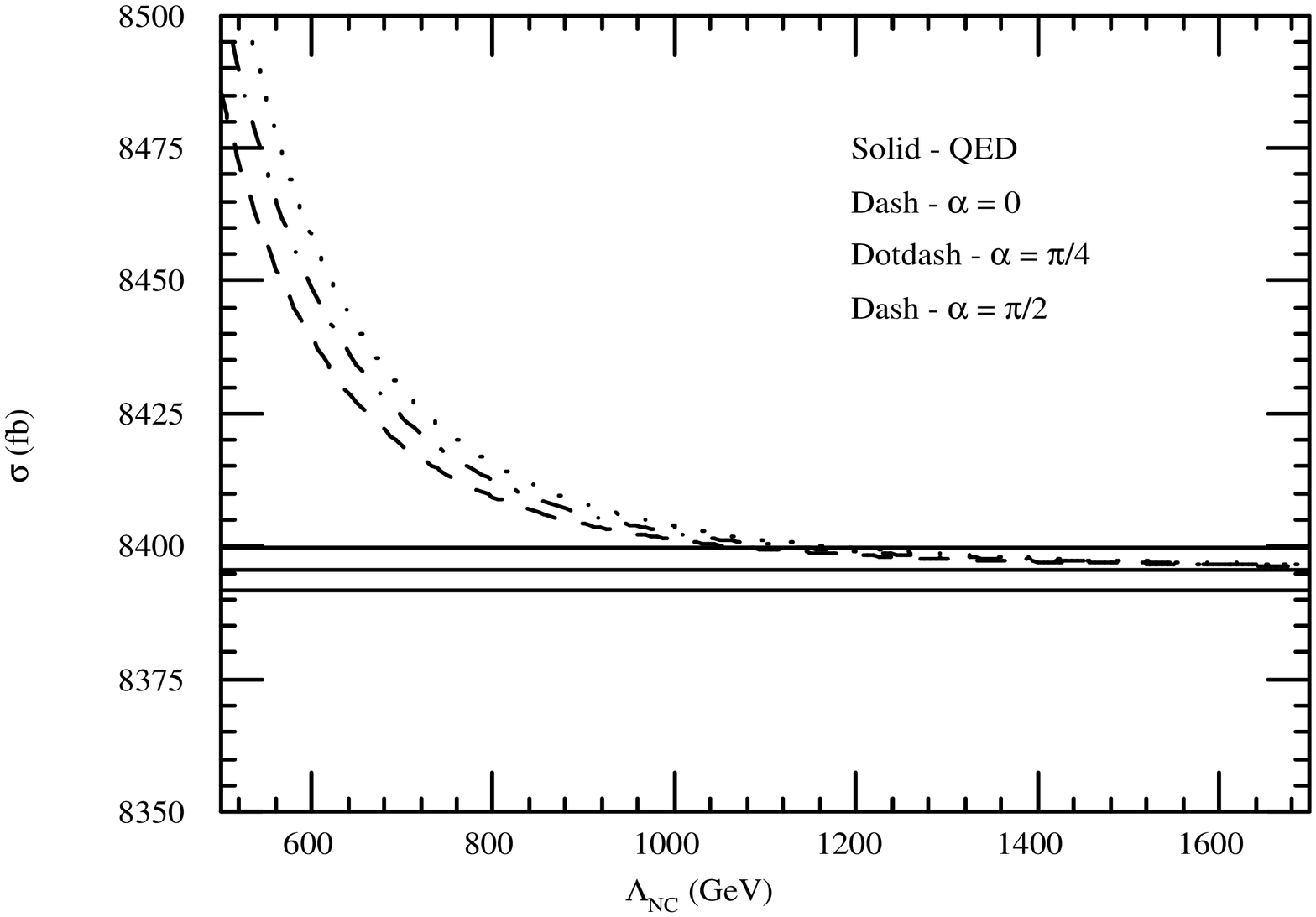} 
}
$\qquad$
\rotatebox{-90}{
\includegraphics[width=5.0cm,clip=true]{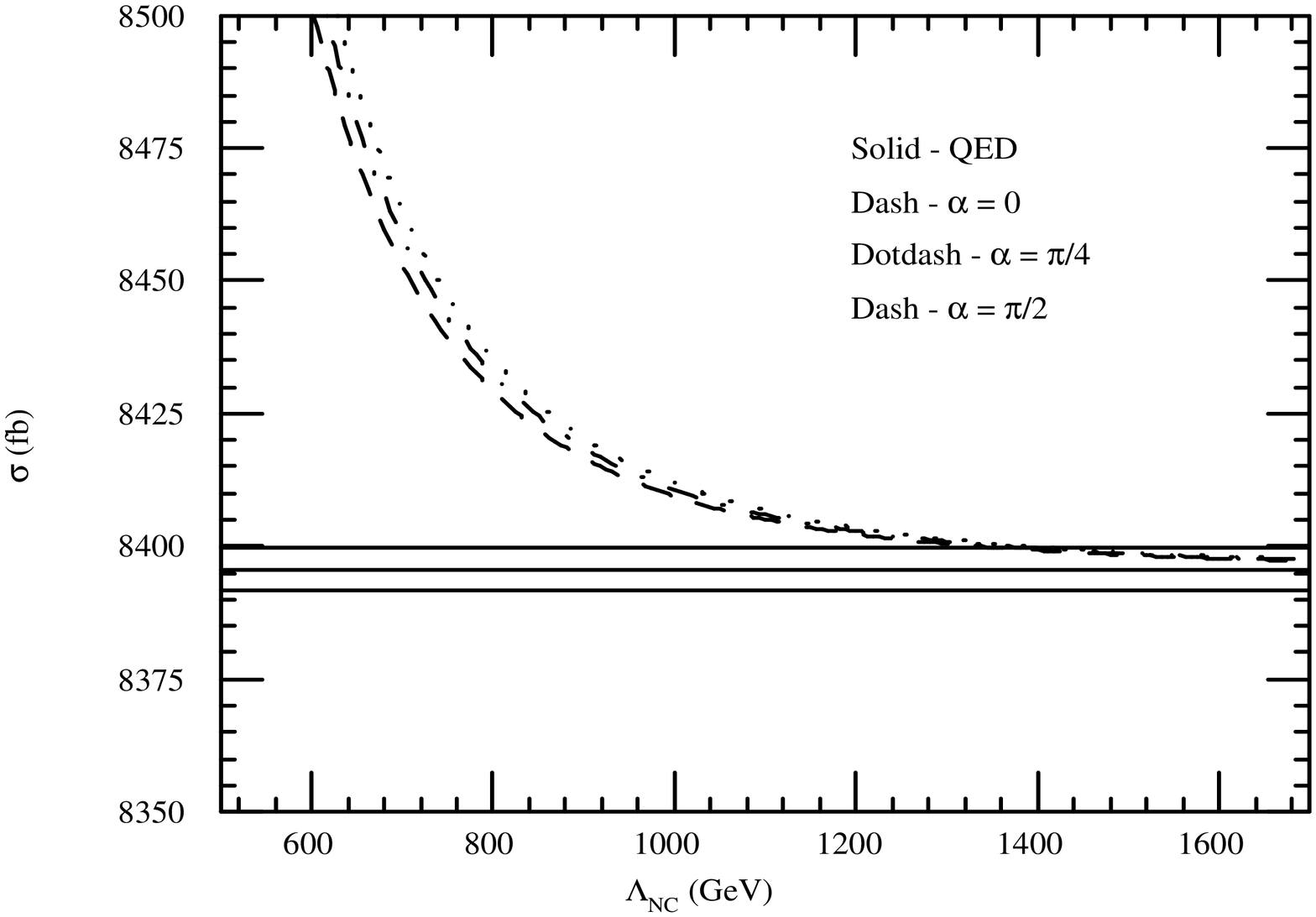} 
}
\caption{
$\sigma$ vs. $\Lambda_{NC}$ for the Compton scattering 
process with $\sqrt{s} = 500$ GeV
for (a) $\gamma=0$ (b) $\gamma=\pi/2$.  
The horizontal band represents the SM cross section 
$\pm$ 1 standard deviation (statistical) error.
}
\end{center}
\end{figure}

Note that there is no $\phi$ dependence for $\alpha=0$ since for this case 
both {\bf E} and {\bf B} are parallel to the beam direction.  In contrast, when
$\alpha=\pi/2$, {\bf E} is perpendicular to the beam direction which is 
reflected in the strong oscillatory behavior in the $\phi$ distribution. 
We find typical limits of $\Lambda_{NC}>(1-2)\sqrt{s}$.

\begin{figure}[t]
\begin{center}
\rotatebox{-90}{
\includegraphics[width=5.0cm,clip=true]{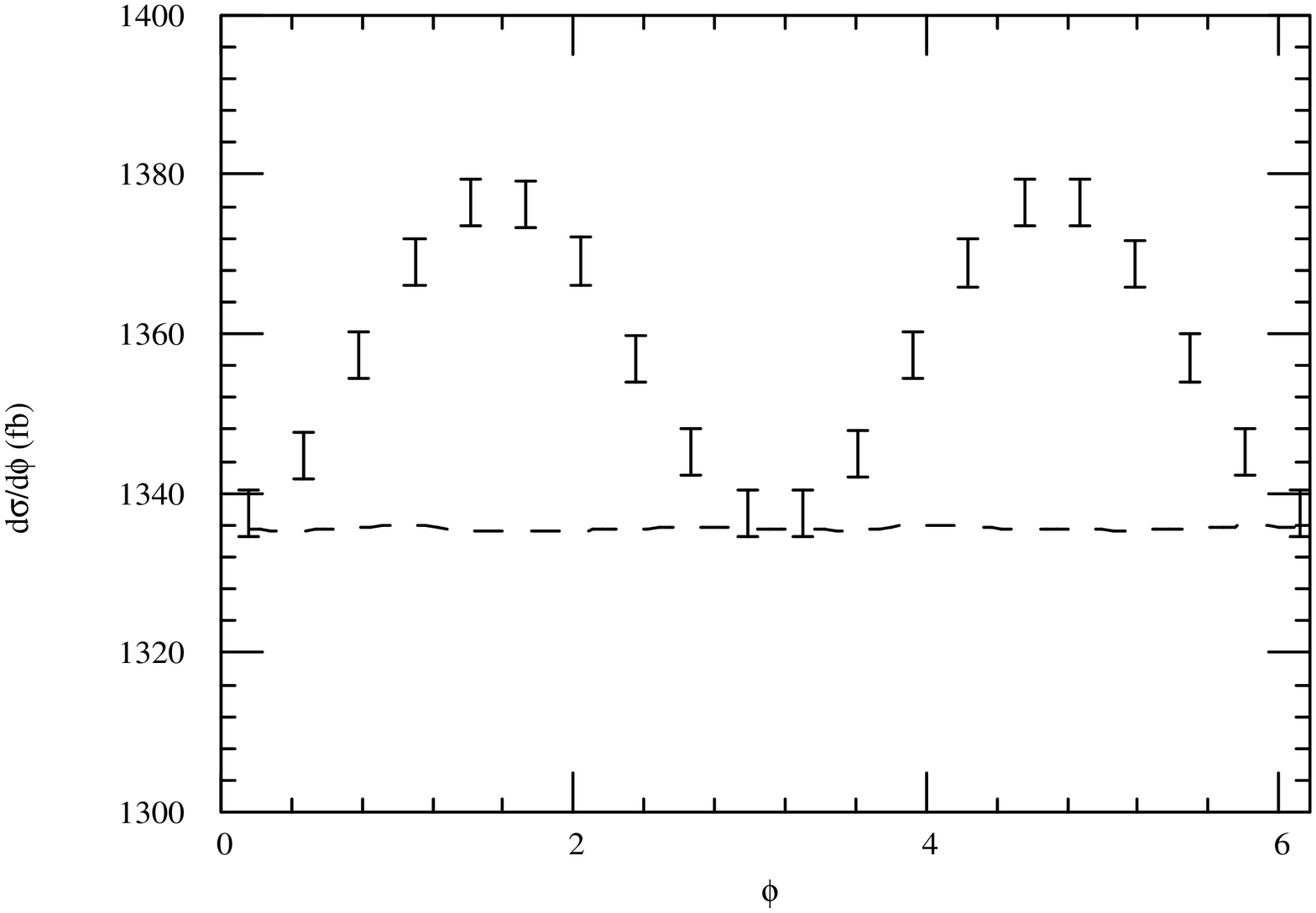} 
}
$\qquad$
\rotatebox{-90}{
\includegraphics[width=5.0cm,clip=true]{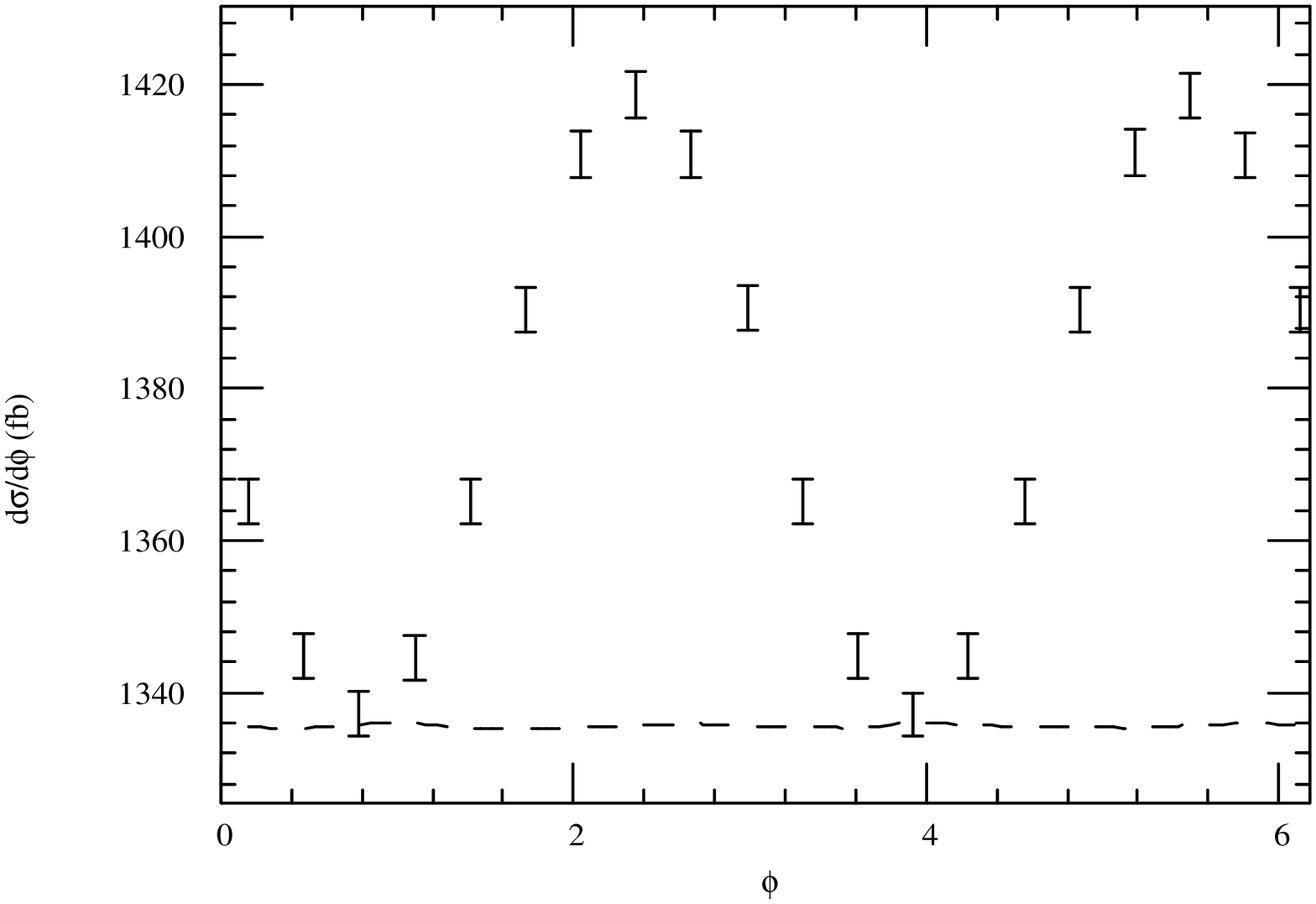} 
}
\caption{
$d\sigma/d\phi$ for the 
Compton scattering process with $\sqrt{s} = 500$ GeV and for
 $\Lambda = 500$ GeV, $\alpha = \pi/2$ and (a) $\gamma = 0$
(b) $\gamma = \pi/2$.  
The dashed curve 
corresponds to the SM angular distribution 
and the points correspond to the NCQED angular distribution including 
1 standard deviation (statistical) error.
}
\end{center}
\end{figure}

\subsection{Other Processes}

In addition to pair creation and Compton scattering other processes 
have been studied.  
Similar results emerge through  momentum dependent electron-photon 
vertices and in the case of pair annihilation, an additional diagram
with the tri-linear photon vertex.  
Similar patterns of azimuthal dependence are found.  
Hewett, Petriollo and Rizzo studied Bhabba scattering, Moller 
scattering, and pair annihilation \cite{hpr}.  
They found that the azimuthal 
dependences could be enhanced by introducing stiffer cuts on the 
angle of the final state particles.  There is a tradeoff between 
reduced statistics with this cut and an enhanced effect.  However, the 
sensitivity is more likely to be limited by systematic errors than by 
statistical errors so the stiffer cuts seem a worthwhile possibility 
to pursue.

In our analysis we chose a specific terrestial 
reference frame to perform our 
calculations.  This is a reasonable approach in a preliminary survey 
of the sensitivity to the NC scale.  However, as we have repeatably 
pointed out, the background fields remain fixed with respect to some 
universal direction.  To perform a proper analysis one would have to 
account for the motion of the experiment with respect to some 
celestial reference frame.  This would need a detailed knowledge of 
not only the location of the experiment with respect to the earth but 
knowledge of when events were collected.  Clearly this is beyond the 
scope of a simple theorist's estimate.  However, along these lines,
Grosse and Liao \cite{gl01} did a
preliminary analysis studying day-night asymetries in $e^+e^-\to HH$ 
production.

\section{Conclusions}

To summarize,  noncommutative field theory is a possibility whose 
phenomenology has received little attention.  In this preliminary 
study we found that lepton pair production and Compton scattering at 
high energy linear colliders are excellent processes to study non-commutative 
QED.  The hallmark signature is the appearance of azimuthally 
dependent cross sections in $2\to 2$ processes.  

The pair production process is only sensitive to space-time NC and is 
therefore insensitive to $\gamma$.  As $\alpha$ increases towards $\pi/2$ the 
deviations from SM decrease towards zero, with $\alpha = \pi/2$ being 
identical to the SM.  On the other hand, the Compton scattering process is 
sensitive to both space-space and space-time NC as parametrized by $\gamma$ 
and $\alpha$.  On the whole, we found that the Compton scattering process is 
superior to lepton pair production in probing NCQED.  Despite significantly 
smaller statistics, the large modification of angular distributions 
leads to higher exclusion limits, well in excess of the 
center of mass energy for all colliders considered.

\section*{Acknowledgements}

This research was supported in part by the Natural Sciences and Engineering 
Research Council of Canada.  

\def \ajp#1#2#3{Am.\ J. Phys.\ {\bf#1}, #2 (#3)}
\def \apny#1#2#3{Ann.\ Phys.\ (N.Y.) {\bf#1}, #2 (#3)}
\def \app#1#2#3{Acta Phys.\ Polonica {\bf#1}, #2 (#3)}
\def \arnps#1#2#3{Ann.\ Rev.\ Nucl.\ Part.\ Sci.\ {\bf#1}, #2 (#3)}
\def \art{and references therein}
\def \cmts#1#2#3{Comments on Nucl.\ Part.\ Phys.\ {\bf#1}, #2 (#3)}
\def \cn{Collaboration}
\def \cp89{{\it CP Violation,} edited by C. Jarlskog (World Scientific,
Singapore, 1989)}
\def \efi{Enrico Fermi Institute Report No.\ }
\def \epjc#1#2#3{Eur.\ Phys.\ J. C {\bf#1}, #2 (#3)}
\def \f79{{\it Proceedings of the 1979 International Symposium on Lepton and
Photon Interactions at High Energies,} Fermilab, August 23-29, 1979, ed. by
T. B. W. Kirk and H. D. I. Abarbanel (Fermi National Accelerator Laboratory,
Batavia, IL, 1979}
\def \hb87{{\it Proceeding of the 1987 International Symposium on Lepton and
Photon Interactions at High Energies,} Hamburg, 1987, ed. by W. Bartel
and R. R\"uckl (Nucl.\ Phys.\ B, Proc.\ Suppl., vol.\ 3) (North-Holland,
Amsterdam, 1988)}
\def \ib{{\it ibid.}~}
\def \ibj#1#2#3{~{\bf#1}, #2 (#3)}
\def \ichep72{{\it Proceedings of the XVI International Conference on High
Energy Physics}, Chicago and Batavia, Illinois, Sept. 6 -- 13, 1972,
edited by J. D. Jackson, A. Roberts, and R. Donaldson (Fermilab, Batavia,
IL, 1972)}
\def \ijmpa#1#2#3{Int.\ J.\ Mod.\ Phys.\ A {\bf#1}, #2 (#3)}
\def \ite{{\it et al.}}
\def \jhep#1#2#3{JHEP {\bf#1}, #2 (#3)}
\def \jpb#1#2#3{J.\ Phys.\ B {\bf#1}, #2 (#3)}
\def \lg{{\it Proceedings of the XIXth International Symposium on
Lepton and Photon Interactions,} Stanford, California, August 9--14 1999,
edited by J. Jaros and M. Peskin (World Scientific, Singapore, 2000)}
\def \lkl87{{\it Selected Topics in Electroweak Interactions} (Proceedings of
the Second Lake Louise Institute on New Frontiers in Particle Physics, 15 --
21 February, 1987), edited by J. M. Cameron \ite~(World Scientific, Singapore,
1987)}
\def \kdvs#1#2#3{{Kong.\ Danske Vid.\ Selsk., Matt-fys.\ Medd.} {\bf #1},
No.\ #2 (#3)}
\def \ky85{{\it Proceedings of the International Symposium on Lepton and
Photon Interactions at High Energy,} Kyoto, Aug.~19-24, 1985, edited by M.
Konuma and K. Takahashi (Kyoto Univ., Kyoto, 1985)}
\def \mpla#1#2#3{Mod.\ Phys.\ Lett.\ A {\bf#1}, #2 (#3)}
\def \nat#1#2#3{Nature {\bf#1}, #2 (#3)}
\def \nc#1#2#3{Nuovo Cim.\ {\bf#1}, #2 (#3)}
\def \nima#1#2#3{Nucl.\ Instr.\ Meth. A {\bf#1}, #2 (#3)}
\def \np#1#2#3{Nucl.\ Phys.\ {\bf#1}, #2 (#3)}
\def \npbps#1#2#3{Nucl.\ Phys.\ B Proc.\ Suppl.\ {\bf#1}, #2 (#3)}
\def \os{XXX International Conference on High Energy Physics, Osaka, Japan,
July 27 -- August 2, 2000}
\def \PDG{Particle Data Group, D. E. Groom \ite, \epjc{15}{1}{2000}}
\def \pisma#1#2#3#4{Pis'ma Zh.\ Eksp.\ Teor.\ Fiz.\ {\bf#1}, #2 (#3) [JETP
Lett.\ {\bf#1}, #4 (#3)]}
\def \pl#1#2#3{Phys.\ Lett.\ {\bf#1}, #2 (#3)}
\def \pla#1#2#3{Phys.\ Lett.\ A {\bf#1}, #2 (#3)}
\def \plb#1#2#3{Phys.\ Lett.\ B {\bf#1}, #2 (#3)}
\def \pr#1#2#3{Phys.\ Rev.\ {\bf#1}, #2 (#3)}
\def \prc#1#2#3{Phys.\ Rev.\ C {\bf#1}, #2 (#3)}
\def \prd#1#2#3{Phys.\ Rev.\ D {\bf#1}, #2 (#3)}
\def \prl#1#2#3{Phys.\ Rev.\ Lett.\ {\bf#1}, #2 (#3)}
\def \prp#1#2#3{Phys.\ Rep.\ {\bf#1}, #2 (#3)}
\def \ptp#1#2#3{Prog.\ Theor.\ Phys.\ {\bf#1}, #2 (#3)}
\def \rmp#1#2#3{Rev.\ Mod.\ Phys.\ {\bf#1}, #2 (#3)}
\def \rp#1{~~~~~\ldots\ldots{\rm rp~}{#1}~~~~~}
\def \rpp#1#2#3{Rep.\ Prog.\ Phys.\ {\bf#1}, #2 (#3)}
\def \sing{{\it Proceedings of the 25th International Conference on High Energy
Physics, Singapore, Aug. 2--8, 1990}, edited by. K. K. Phua and Y. Yamaguchi
(Southeast Asia Physics Association, 1991)}
\def \slc87{{\it Proceedings of the Salt Lake City Meeting} (Division of
Particles and Fields, American Physical Society, Salt Lake City, Utah, 1987),
ed. by C. DeTar and J. S. Ball (World Scientific, Singapore, 1987)}
\def \slac89{{\it Proceedings of the XIVth International Symposium on
Lepton and Photon Interactions,} Stanford, California, 1989, edited by M.
Riordan (World Scientific, Singapore, 1990)}
\def \smass82{{\it Proceedings of the 1982 DPF Summer Study on Elementary
Particle Physics and Future Facilities}, Snowmass, Colorado, edited by R.
Donaldson, R. Gustafson, and F. Paige (World Scientific, Singapore, 1982)}
\def \smass90{{\it Research Directions for the Decade} (Proceedings of the
1990 Summer Study on High Energy Physics, June 25--July 13, Snowmass, Colorado),
edited by E. L. Berger (World Scientific, Singapore, 1992)}
\def \tasi{{\it Testing the Standard Model} (Proceedings of the 1990
Theoretical Advanced Study Institute in Elementary Particle Physics, Boulder,
Colorado, 3--27 June, 1990), edited by M. Cveti\v{c} and P. Langacker
(World Scientific, Singapore, 1991)}
\def \yaf#1#2#3#4{Yad.\ Fiz.\ {\bf#1}, #2 (#3) [Sov.\ J.\ Nucl.\ Phys.\
{\bf #1}, #4 (#3)]}
\def \zhetf#1#2#3#4#5#6{Zh.\ Eksp.\ Teor.\ Fiz.\ {\bf #1}, #2 (#3) [Sov.\
Phys.\ - JETP {\bf #4}, #5 (#6)]}
\def \zpc#1#2#3{Zeit.\ Phys.\ C {\bf#1}, #2 (#3)}
\def \zpd#1#2#3{Zeit.\ Phys.\ D {\bf#1}, #2 (#3)}

\end{document}